\documentstyle[aps,amssymb,multicol]{revtex}
\newcommand {\be}{\begin {equation}}
\newcommand {\ee}{\end {equation}}

\begin {document}

\title {The interaction of surface acoustic waves with an array of quantum
        wires}

\author {Michal Rokni and Y. Levinson}
\address {Department of Condensed Matter Physics,
 The Weizmann Institute of Science, Rehovot 76100, Israel}

\date {February 21, 1999}

\maketitle

\begin {abstract}

We describe the interaction of surface acoustic waves with electrons in
an array of quantum wires, patterned out of a two dimensional electron
gas. Two specific geometries are considered, in which the surface
acoustic wave travels parallel, or perpendicular to the wires. Although
we assume the electron wave functions in different wires do not
overlap, the screening of the phonon potential by the electrons in the
wire array is a collective phenomenon. It is shown that the surface
acoustic wave absorption cannot be described via the ac conductivity in
the usual manner. We derive an integral equation for the dielectric
function of the electrons in the quantum wire system, and solve it in
the narrow wire approximation. Using the dielectric function we find
the absorption and the change in velocity of the surface acoustic
waves.
\end {abstract}

\vspace{1cm}

\begin{multicols}{2}

\section {Introduction}
Experiments which include interactions of surface acoustic waves
(SAW's) with electrons in a two dimensional electron gas (2DEG) are of
great interest \cite {WKW86,WSW89}. Unlike conductivity measurements
in which a current is driven through the 2DEG and the voltage is
measured, where only the conductivity at zero wave vector can be
probed, SAW measurements allow one to probe the finite wave vector
conductivity. Experiments that have been carried out in the fractional
quantum Hall regime near a filling factor $ \nu = 1/2 $ \cite
{WPR90,WRP93,WRW93}, have attracted great attention, since they
strongly support the composite fermion approach to the compressible
state at $ \nu = 1/2 $.

Experimental work done in modulated 2DEG \cite {WWP97} triggered theoretical
works on SAW's in such systems. In Ref.\ \cite {LEM98} Weiss oscillations
in SAW propagation were considered, while composite fermions
in modulated structures were treated in \cite {MWL98,OSH98}.

This work was motivated by an experiment \cite {NBB96}, in which the
transmission of a SAW by an array of parallel quantum wires (QWR's) was
measured, as a function of a strong magnetic field (that tunes the electron
subbands in the QWR's relative to the Fermi energy). The measurements
were performed in two geometries with the SAW traveling parallel or
perpendicular to the QWR's. In the latter case a structure that corresponds
to the crossing of the Fermi level by the bottom of the subbands was observed,
while in the former case no such structure was seen.

In this work we will calculate the absorption and the change in velocity
of a SAW that propagates either parallel or perpendicular to
a QWR array. In section \ref {formulation} we describe the model that we
consider.  We then show why the ``usual'' - ``classical'' approach
cannot be used to describe the SAW attenuation by the electrons in the
QWR array. When the Boltzmann equation is used in order to calculate
the phonon relaxation rate due to electron transitions, one finds that
the SAW absorption is negligible, since the phase space available for
these transitions in one dimension is highly restricted by energy and
momentum conservation. The approach in which the SAW absorption is
given in terms of the dc conductivity also fails since in this system
the dc conductivity is zero. When the SAW travels parallel to the wires
the dc conductivity is zero due to localization, while in the
perpendicular case it is zero since the electron wave functions in
different wires do not overlap.

In section \ref {dielectric} we will define what the
dielectric matrix is, and explain how it is related to the SAW
absorption and change in velocity. We then find an equation for the
inverse dielectric matrix in terms of the polarization (that is also
defined in this section). In section \ref {wf} we describe the random
potential that we consider, and the wave functions that are related to it.
We quote the result for a specific average over the wave functions, that we
shall encounter in the calculation of the polarization, and discuss the
conditions for its validity. The calculations of the polarization and the
inverse dielectric matrix are presented in section \ref {calculation}.
The results will be discussed in section \ref {discussion}.

\section {Formulation of the problem}
\label {formulation}

The geometry of our system is as follows: There are $ N $ wires of
length $ L_{y} $, and width $ w $, in a periodic array of period $ d $
($ L_{x} = N d $). The coordinate along the wires is $ y $, and the
coordinate perpendicular to the wires is $ x $. The electrons in a
single wire are described by a wave function $ \psi_{n \alpha}(x,y) $,
that corresponds to the eigenenergy $ E_{n \alpha} $. We assume that
the electron wave functions in different wires do not overlap. SAW's of
wave vector $ {\bf q} = (q_{x},q_{y}) $, and frequency $ \omega_{q} =
v_{S}q $ (where $ q = |{\bf q}| $), interact with the electrons in this
array. Since the
effect that is seen in the experiment is clearly connected with the
crossing of the Fermi level by the electron energy subbands, we
consider the magnetic field only as a mechanism for moving the position
of the bottom of the subbands, and we do not consider the manner in
which it affects the wave functions of the electrons.

When the electrons are free i.e., they do not interact and they are not
under the influence of a random potential, the electron wave functions
and energies are separable, and can be written as
$ \psi_{n, k}(x,y) = \phi_{n}(x) e^{i k y}/\sqrt {L_{y}} $, and
$ E_{n, k} = E_{n} + \epsilon_{k} $.
Here $ E_{n} $ are the energy subbands created by the constriction in the
$ x $ direction, and $ \phi_{n}(x) $ are the corresponding wave functions.
The kinetic energy  is given by $ \epsilon_{k} = k^2/2 m^* $ where $ m^* $
is the effective mass.

Let us first show that the absorption of the SAW cannot be described
using a simple Boltzmann formalism. If one were to use the Boltzmann equation
and the ``free'' electron states in order to calculate the phonon decay rate
$ 1/\tau $, one would obtain
\begin{eqnarray}
  \label {rate-BE}
  \frac {1}{\tau} & = & \frac
  {1}{d} \sum_{n} |M_{n}(q_{x})|^2 \int dk \,
  \delta(\epsilon_{k} - \epsilon_{k-q_{y}} - \omega_{q})
  \nonumber \\ & \times & \left[ f(E_{n,k-q_{y}}) - f(E_{n,k})\right].
\end{eqnarray}
The phonon energy $ \omega_{q} $ is much smaller than the subband difference
$ \Delta E_{n} $, therefore no intersubband transitions occur due to the
electron-phonon interaction.  The matrix element of the electron-phonon
interaction is given by
$ M_{n}(q_{x}) = M^{o} \int_{0}^{w} dx \, |\phi_{n}(x)|^2 e^{iq_{x} x} $,
where
$ M^{o} = 4 \pi \beta e (\hbar / a \rho v_{S})^{1/2} C $. The constant
$ \beta $ describes the piezoelectric coupling, $\rho $ is the mass density of
the material, and $ a $ and $ C $ are numerical factors that depend on the
SAW: on the direction of its propagation, its velocity, and the elastic
constants of the material through which it propagates \cite {KLE96}. The
constant $ a $ is of the order of $ 1 $ and the constant $ C $
is of the order of $ 0.1 $. The function $ f(E_{n,k}) $ is the
electron distribution function, and at equilibrium it is given by the
Fermi distribution function, $ f_{T}(E_{n,k}) $.

The delta function that appears in expression (\ref {rate-BE}) for $ 1/\tau $
stands for energy and momentum conservation. When the phonon travels parallel
to the wires i.e., when $ {\bf q} \parallel \hat {\bf y} $, then one obtains
from Eq.\ (\ref {rate-BE})
\begin{eqnarray}
  \label {rate-BE-par}
  \frac {1}{\tau} & = & \frac {m^*}{q d}
  \sum_{n} |M_{n}(0)|^2 \nonumber \\ & \times &
  \left[ f_{T}(E_{n} + \epsilon_{m^* v_{S} - q/2}) -
         f_{T}(E_{n} + \epsilon_{m^* v_{S} + q/2}) \right].
\end{eqnarray}
If the temperature $ T \gg \omega_{q} $, SAW absorption occurs when
$ \epsilon_{F_{n}} \equiv \epsilon_{F} - E_{n} $ (the Fermi energy
measured from the bottom of the $ n^{\rm th} $ subband) is within $ T $ of
$ m^* v_{S}^2 $. If $ T \ll \omega_{q} $, SAW absorption occurs only when
$ \epsilon_{F_{n}} $ is within $ \omega_{q} $ of $ m^* v_{S}^2 $.
For both temperature regions the peaks are expected to be of square shape.
For $ T = 1.3 \, {\rm K} $, and $ \omega_{q} = 300 \, {\rm neV} $, the
temperature and phonon frequency at which the experiment
\cite {NBB96} was carried out, one would expect peaks of the width of
$ \Delta B = 0.06 \, {\rm  T} $, while in practice no peaks were seen.

If $ {\bf q} \parallel \hat {\bf x} $, Eq.\ (\ref {rate-BE}) leads
to,
\begin{eqnarray}
  \label {rate-BE-per}
  \frac {1}{\tau} & = & \frac {1}{d}
  \sum_{n} |M_{n}(q)|^2 \, \delta (\omega_{q})
  \nonumber \\
  & \times & \int dk \,
  \left[ f_{T}(E_{n,k-q_{y}}) - f_{T}(E_{n,k})\right] = 0,
\end{eqnarray}
regardless of the temperature.
Thus, in this case there is no absorption at all, contrary to the
absorption peaks that were observed in the experiment, that were of the width
of $ \Delta B \approx 1 \, {\rm  T} $, and were not of a square shape.

From the two cases described above it is clear that if one wants to
explain a finite SAW absorption in both the parallel and the
perpendicular cases, as is seen for example in \cite {NBB96}, a
nonzero temperature does not suffice, and one has to take into account
either inelastic, or elastic scattering. These will introduce a level
broadening that will relax the energy and the momentum conservation
requirements. We shall consider only elastic scattering, specifically
that due to a random potential. In addition we assume zero temperature
for simplicity.

In the usual approach the SAW attenuation is given in terms of the dc
conductivity by
\be
  \label {Gamma-dc}
  \Gamma = -\frac {\alpha^2}{2}
  \frac {\sigma_{\rm dc}/\sigma_{\rm M}}
        {1 + (\sigma_{\rm dc}/\sigma_{\rm M})^2},
\ee
where $ \alpha $ is the piezoelectric coupling constant,
$ \sigma_{\rm M} = \kappa_{o} \omega_{q} / 2 \pi q $ is the Maxwell
conductivity, and $ \kappa_{o} $ is the dielectric constant of the material.
We cannot use (\ref {Gamma-dc}) since in our case $ \sigma_{\rm dc} = 0 $.
For $ {\bf q} \parallel $ QWR's the dc conductivity is zero since, as is
well known, the states in a one dimensional system in the presence of
disorder are localized (see for example the review \cite {LeR85} and
references therein).
In the case of $ {\bf q} \perp $ QWR's the dc conductivity is zero since the
functions in different wires do not overlap.

Although the electron wave functions in different wires do not overlap,
so that the absorption is additive, the electrons in different wires interact
via the Coulomb potential, giving rise to a strong collective screening effect.

\section {The dielectric function}
\label {dielectric}
A function that will describe both the absorption of the SAW by the electrons
in the QWR array, and the strong screening, is the dielectric function.
When a SAW of frequency $ \omega $ travels through the sample, the potential
exerted by it $ \varphi_{\omega} $, is screened by the electrons. The
screened potential $ \varphi^{\rm scr}_{\omega} $ is related to
$ \varphi_{\omega} $ through the inverse dielectric function
$ \epsilon^{-1}_{\omega} $
\be
  \label {varphi-scr}
  \varphi^{\rm scr}_{\omega}({\bf r}) = \int d{\bf r}' \,
  \epsilon^{-1}_{\omega}({\bf r}, {\bf r}') \, \varphi_{\omega}({\bf r}'),
\ee
therefore the absorption of the SAW by the electrons in the wire array is
related to the inverse dielectric function.

If the 2DEG is periodically patterned in the $ x $
direction with a period $ d $, and homogeneous in the $ y $ direction,
then the nonzero Fourier components $ \epsilon^{-1}_{\omega}
({\bf q',q}'') $
have $ q'_{x} = q_{x} + 2 \pi s'/d \equiv q_{s'} $,
     $ q''_{x} = q_{x} + 2 \pi s''/d \equiv q_{s''}$
($ s', s'' $ are integers), and
$ q'_{y} = q''_{y} \equiv q_{y} $.
These components can be written in the form of a matrix
$ (\epsilon^{-1})_{s',s''}({\bf q},\omega) $ \cite {LEM98}, where
$ {\bf q} = (q_{x}, q_{y}) $ (as we will explain
below). We shall denote $ (\epsilon^{-1})_{s',s''} $ by
$ \epsilon^{-1}_{s',s''} $ from now on in order to simplify the notations.
In terms of these components the attenuation per unit length, and the
change of the velocity of the SAW, are given by \cite {LEM98}
\be
  \label {Gamma}
  \Gamma = - q \frac {\alpha^2}{2} \,
  {\rm Im} \, \epsilon^{-1}_{0,0}({\bf q}, \omega_{q}),
\ee and
\be
  \label {dv/v}
  \frac {\Delta v}{v} = \frac {\alpha^2}{2} \,
  {\rm Re} \left[ \epsilon^{-1}_{0,0}({\bf q}, \omega_{q}) - 1 \right].
\ee
Our goal is therefore be to find
$ \epsilon^{-1}_{0,0}({\bf q}, \omega_{q}) $.

The dielectric function is given by the following equation (for example
see \cite {Mah})
\end{multicols}
\widetext

\noindent
\begin{picture}(3.375,0)
  \put(0,0){\line(1,0){3.375}}
  \put(3.375,0){\line(0,1){0.08}}
\end{picture}
\begin {eqnarray}
  \label {eps-space}
 \epsilon^{-1}(x,x';y-y';\omega_{q}) & = &
  \delta (x-x') \delta (y-y')
   \nonumber \\ & + &
  \int \!\! \int_{0}^{L_{x}} \!\! dx_{1} \, dx_{2}
  \int \!\! \int_{0}^{L_{y}} \!\! dy_{1} \, dy_{2}
  V(x-x_{1};y-y_{1}) \Pi(x_{1},x_{2};y_{1}-y_{2};\omega_{q}) \,
  \epsilon^{-1}(x_{2},x';y_{2}-y';\omega_{q}) ,
\end {eqnarray}
where $ V ({\bf r}-{\bf r}') = e^2/(\kappa_{o} |{\bf r} - {\bf r}'|) $ is
the bare Coulomb interaction between two electrons situated at $ {\bf r} $
and $ {\bf r}' $, in a material of dielectric constant $ \kappa_{o} $. The
retarded polarization $ \Pi $ is given by
\begin {eqnarray}
  \label {Pi-space}
  \lefteqn {\Pi (x_{1},x_{2};y_{1}-y_{2};\omega_{q}) =}
   \nonumber
  \\ & & - \frac {i}{2} \,
  \Biggl \langle
    \int \! \frac {d \epsilon}{2 \pi}
    \Bigl[ G^{\rm s}(x_{1},x_{2};y_{1} \! - \! y_{2};\epsilon) \,
           G^{\rm a}(x_{2},x_{1};y_{2} \! - \! y_{1};\epsilon \! - \! \omega_{q}) +
           G^{\rm r}(x_{1},x_{2};y_{1} \! - \! y_{2};\epsilon) \,
           G^{\rm s}(x_{2},x_{1};y_{2} \! - \! y_{1};\epsilon \! - \! \omega_{q}) \Bigr]
  \Biggr \rangle_{U},
\end {eqnarray}
\hfill
\begin{picture}(3.375,0)
  \put(0,0){\line(1,0){3.375}}
  \put(0,0){\line(0,-1){0.08}}
\end{picture}

\begin{multicols}{2}
\noindent where $ G^{\rm s},G^{\rm a}, $ and $ G^{\rm r} $ are the
statistical, advanced, and retarded one particle electron Green
functions (for definitions of these functions see \cite {LiP}).
The electrons in the QWR array are at equilibrium, therefore
expression (\ref {Pi-space}) is simplified by the substitution of
$ G^{\rm s}({\bf r},{\bf r}';\epsilon) = 2 i (1 - 2 f(\epsilon))
\, {\rm Im} \, G^{\rm r}({\bf r},{\bf r}';\epsilon) $, where $ f $
is the Fermi distribution. The angular brackets $ \langle \cdots
\rangle_{U} $ represent averaging over all impurity
configurations.

The dielectric function and the polarization are
considered as quantities that are averaged over all impurity configurations
(correlations between the two are neglected so that each is averaged
separately), therefore they are translationally invariant in the $ y $
direction. Since we assume the electron wave functions in different wires do
not overlap, the two variables of $ \Pi $ in the $ x $ direction are restricted
to the same wire. This is not the case with $ \epsilon^{-1} $, where two
electrons in different wires can interact via the Coulomb interaction.

We now wish to Fourier transform the equation above. Due to the homogeneity in
the $ y $ coordinate, the equation will be diagonal in $ q_{y} $.
Using the following definitions
\begin{eqnarray}
  \label {eps-q}
  \lefteqn {\epsilon^{-1}(q_{x}, q'_{x};q_{y};\omega_{q})  =
  \frac {1}{L_{x}}
  \int \! \! \int_{0}^{L_{x}} dx \, dx' \int_{0}^{L_{y}} dy}
  \nonumber \\ & & \,\,\,\,\,\,\,\,\,\,\,\,\,\,\,\,\,\, \times
  e^{-i q_{x} x + i q'_{x}x' - iq_{y} y} \,
  \epsilon^{-1}(x,x';y;\omega_{q}),
\end{eqnarray}
\be
  \label {V-q}
  V({\bf r}) = \frac {1}{L_{x} L_{y}} \sum_{\bf q}
               e^{i {\bf q} \cdot {\bf r}} \, V({\bf q}),
\ee
where $ V({\bf q}) = 2 \pi e^2/(\kappa_{o} q) $, and
\begin{eqnarray}
  \label {Pi-q}
  \lefteqn {\Pi (x, x';y;\omega_{q}) =} \nonumber \\ & &
  \frac {1}{L_{x}^2 L_{y}} \sum_{q_{x},q_{x}',q_{y}}
  e^{i q_{x} x - i q_{x}' x' + i q_{y} y}
  \Pi^{N}(q_{x}, q_{x}';q_{y};\omega_{q}),
\end{eqnarray}
we Fourier transform Eq.\ (\ref {eps-space}) and obtain
\begin{eqnarray}
  \label {Eq-eps-q}
  \lefteqn {\epsilon^{-1}(q_{x}, q_{x}'; q_{y};\omega_{q}) = \delta_{q_{x}, q'_{x}}}
  \\ \nonumber & & +
  V(q_{x},q_{y}) \frac {1}{L_{x}} \sum_{q''_{x}}
  \Pi^{N} (q_{x}, q''_{x};q_{y};\omega_{q})
  \epsilon^{-1}(q''_{x}, q'_{x};q_{y};\omega_{q}).
\end{eqnarray}

Due to the periodicity of the array,
$ \epsilon^{-1}(x_{1},x_{2};y;\omega_{q}) $ and
$ \Pi(x_{1},x_{2};y;\omega_{q}) $ remain unaltered when their two $ x $
variables, $ x_{1} $ and $ x_{2} $, are replaced by $ x_{1} + nd $ and
$ x_{2} + nd $ ($ n $ is any integer).
Therefore the $ x $ components of the two $ {\bf q} $ variables of
$ \epsilon^{-1}(q_{1x}, q_{2x};q_{y};\omega_{q}) $ and
$ \Pi^{N}(q_{1x}, q_{2x};q_{y};\omega_{q}) $ will differ by $ 2 \pi s/d $,
where $ s $ is an integer. This is due to the fact that in an unmodulated
structure the screened field will have the same wave vector as the unscreened
field, while in a modulated structure the screened field will have all the
Umklapp wave vectors that are related to the inverse lattice of the modulation,
as well. Thus if we define
$ q_{s} = q_{x} + 2 \pi s/d $, where $ q_{x} $ is the $ x $ component of the
phonon wave vector, then
\be
  \label {eps-sq}
  \epsilon^{-1}(q_{1x}, q_{2x}; q_{y}; \omega_{q}) =
  \epsilon^{-1}(q_{s}, q_{s'}; q_{y};\omega_{q} )
  \equiv
  \epsilon^{-1}_{s,s'}({\bf q},\omega_{q}).
\ee

Rewriting Eq.\ (\ref {Eq-eps-q}) in terms of the new $ s $ variables we obtain
\begin{eqnarray}
  \label {Eq-eps-sq}
  \epsilon^{-1}_{s,s'}({\bf q},\omega_{q}) & = & \delta_{s, s'}
  \nonumber \\ & + &
  V_{s}({\bf q}) \frac {1}{L_{x}} \sum_{s''}
  \Pi^{N}_{s,s''} ({\bf q},\omega_{q})
  \epsilon^{-1}_{s'',s'}({\bf q},\omega_{q}),
\end{eqnarray}
where $ V_{s}({\bf q}) = 2 \pi e^2 /[\kappa_{o} (q_{s}^2 + q_{y}^2)^{1/2}] $.
Note that in our case $ q_{y} $ that appears in the equation is the $ y $
component of the phonon momentum, while $ q_{x} $, the basis for all $ q_{s} $,
is the $ x $ component of the phonon momentum. Thus $ {\bf q} $ is the
phonon momentum.

Reversing the relation given by (\ref {Pi-q}), and taking into account the
fact that the $ x $ coordinates of $ \Pi $ are restricted to the same wire
we find that
\be
  \label {PiN-sq}
  \Pi^{N}_{s,s''} ({\bf q},\omega_{q}) =
  \sum_{m=0}^{N} \Pi_{s,s''}^{m} ({\bf q},\omega_{q}),
\ee
where
\begin{eqnarray}
  \label {Pim-sq}
  \Pi_{s,s''}^{m} ({\bf q},\omega_{q}) & = &
  \int \!\! \int_{md}^{md+w} dx \, dx' \int_{0}^{L_{y}} dy
  \nonumber \\ & \times &
  e^{-i q_{s} x + i q_{s'} x' - i q_{y} y} \, \Pi (x,x';y,\omega_{q}),
\end{eqnarray}
is the polarization of the $ m^{\rm th} $ wire. Since all the wires are
identical, and there is no overlap between the electron wave functions,
the polarization of all wires must be the same and the suffix $ m $ can be
dropped
\be
  \label {Pi-sq}
  \Pi^{N}_{s,s''} ({\bf q},\omega_{q}) =
  N \Pi_{s,s''} ({\bf q},\omega_{q}).
\ee
This leads to the final form of the equation for the inverse dielectric matrix
\be
  \label {Eq-eps-sq1}
  \epsilon^{-1}_{s,s'}({\bf q},\omega_{q}) =
  \delta_{s,s'} +
  \frac{1}{d} V_{s} ({\bf q}) \sum_{s''} \Pi_{s,s''}({\bf q}, \omega_{q})
  \epsilon^{-1}_{s'',s'}({\bf q}, \omega_{q}).
\ee

In order to solve Eq.\ (\ref {Eq-eps-sq1}) for $
\epsilon^{-1}_{s,s'} $ we must find the polarization $ \Pi_{s,s''}
$. It can be written in terms of the electron wave functions and
energies in the following form

\begin{picture}(0,0.2)
\end{picture}

\end{multicols}
\widetext

\noindent
\begin{picture}(3.375,0)
  \put(0,0){\line(1,0){3.375}}
  \put(3.375,0){\line(0,1){0.08}}
\end{picture}
\be
  \label{Pi-sq-wf}
  \Pi_{s,s'}({\bf q},\omega_{q}) = \frac {1}{L_{y}}
  \Biggl \langle \sum_{n,n';\alpha,\alpha'}
    \langle n' \alpha' |e^{-i q_{s} x - i q_{y} y}| n \alpha \rangle
    \langle n \alpha |e^{i q_{s'} x + i q_{y} y}| n' \alpha' \rangle
    \frac {f(E_{n' \alpha'}) - f(E_{n \alpha})}
          {\omega_{q} - (E_{n \alpha} - E_{n' \alpha'}) + i \delta}
  \Biggr \rangle_{U}.
\ee \hfill
\begin{picture}(3.375,0)(0,0)
  \put(0,0){\line(1,0){3.375}}
  \put(0,0){\line(0,-1){0.08}}
\end{picture}

\begin{multicols}{2}

\begin{picture}(0,0)
\end{picture}
\vspace{-0.5cm}

\section {The random potential and the wave functions}
\label {wf}

We will assume that the random potential responsible for the states
$ \psi_{n \alpha} $ can be written as a sum of the random potential
$ U_{0}(x,y) $ for the unpatterned 2DEG (with amplitude
$ \langle U_{0}^2 \rangle $ and correlation length $ \Lambda_{0} $),
and an effective potential $ V_{n}(y) $ which describes the scattering by the
QWR's boundaries. When the width of the QWR $ w  \ll \Lambda_{0} $, the total
potential within the QWR can be written as $ U(x,y)=U_{n}(y) + xW(y) $, where
$U_{n}(y)=V_{n}(y)+U_{0}(0,y)$, and $ W(y)= \partial U_{0}(0,y)/\partial x $.
We shall assume that $ U_{0}(0,y) $ is a Gaussian  correlated random variable
with $ \left \langle U_{0}(0,y)  U_{0}(0,y') \right \rangle =
       \langle U_{0}^2 \rangle
       \exp {\left[(y -y')^2/ 2 \Lambda_{0}^2 \right]}$,
therefore $ W(y) $ and $ U_{0}(0,y) $ are uncorrelated. In addition,
$ V_{n}(y) $ and $ W(y) $ are uncorrelated (since they stem from different
unrelated physical processes). We thus conclude that $ W(y) $ and $ U_{n}(y) $
are uncorrelated.

When the term containing $ W $ can be neglected, the wave
functions and the energies are separable i.e., $ \psi _{n
\alpha}(x,y) = \phi_{n}(x) \chi_{n \alpha}(y) $, and $ E_{n
\alpha}^{0} = E_{n} + \epsilon_{n \alpha} $. Here $ \phi_{n} $ and
$ E_{n} $ are defined by the confining potential, while $ \chi_{n
\alpha} $ and $ \epsilon_{n \alpha} $ are defined by the potential
$ U_{n} $. Thus only $ \chi_{n \alpha} $ and $ \epsilon_{n \alpha}
$ are random. This approximation will suffice in the case of $
{\bf q} \parallel $ QWR's, but for $ {\bf q} \perp $ QWR's a
higher order approximation that includes the linear term in $ x W
$ is needed
\begin{eqnarray}
  \label {ho-wf}
  \lefteqn {\psi_{n \alpha}(x,y) =
  \phi_{n}(x) \chi_{n \alpha}(y) } \nonumber \\ & + &
  \sum_{n' \alpha'}
  \frac {x_{n' n} W_{n' \alpha';n \alpha}}
        {E^{0}_{n \alpha}- E^{0}_{n' \alpha'}} \,
  \phi_{n'}(x) \chi_{n' \alpha'}(y),
\end{eqnarray}
where
\be
   x_{n' n} = \int_{0}^{w} dx \, \phi_{n'}^{*}(x) \, x \,
  \phi_{n}(x),
\ee
\be
  W_{n' \alpha';n \alpha} = \int_{0}^{L_{y}} dy \,
  \chi_{n' \alpha'}^{*}(y) W(y) \chi_{n \alpha}(y).
\ee
Under the assumption of symmetric quantum wires, the diagonal matrix
elements of $ x $ are $ x_{n,n} = 0 $.

We assume the wires to be narrow $ w \ll d $, and approximate the following
matrix elements by
\be
  \label {exp-me}
  \int_{0}^{w} \!\! dx \, \phi_{n}^*(x) e^{i q_{s} x}
                     \phi_{n'}(x) = \left\{ \!\!
  \begin {array}{ll}
    \delta_{n,n'} + iq_{s}  x_{n n'} & {\rm if} \\ & s \leq d/2 \pi w,
    \\ \\
    0 & {\rm otherwise}.
  \end {array} \right.
\ee
The justification for this approximation is that for $ s \ll d/2 \pi w $ we can
expand the exponent, while for $ s \gg d/2 \pi w $ the matrix elements are
exponentially small.

In our calculations we shall encounter the following average over electron
wave functions and energies
\end{multicols}
\widetext
\be
  \label {average}
  \Biggl \langle \sum_{\alpha,\alpha'}
  \delta (\epsilon_{n \alpha} - \epsilon)
  \delta (\epsilon_{n \alpha'} - \epsilon')
  \chi^{*}_{n \alpha'}(y) \chi_{n \alpha}(y)
  \chi_{n \alpha'}(y') \chi^{*}_{n \alpha}(y')
  \Biggr \rangle_{U}.
\ee
This average has been calculated in \cite {GDP83b}, where the
averages of electron wave functions in one dimensional structures,
over ensembles of random impurities, are calculated. The following
result was obtained in \cite {GDP83b}
\begin {eqnarray}
  \label {F}
  F^{(1)}_{n}(\xi) & \equiv &
  \frac {\sqrt {\epsilon \, \epsilon'}}{\epsilon_{Fn}}
  \left \langle \sum_{\alpha,\alpha'}
  \delta (\epsilon_{n \alpha} - \epsilon)
  \delta (\epsilon_{n \alpha'} - \epsilon')
  \chi^{*}_{n \alpha'}(y) \chi_{n \alpha}(y)
  \chi_{n \alpha'}(y') \chi^{*}_{n \alpha}(y')
  \right \rangle_{U} \nonumber \\
  & = &
  \left\{
    \begin {array}{lll}
      \frac {2}{3} \nu^2(\epsilon_{Fn}) =
      \frac {2}{3} (\pi v_{Fn})^{-2} & {\rm if} \;\;\;\;
      k_{Fn}^{-1} \ll \xi \ll l_{n} & \\ \\
      \frac {\pi^{7/2}}{16} \nu^2(\epsilon_{Fn})
      \left( l_{n}/\xi \right)^{3/2}
      \exp {- \left( \xi/ 4 l_{n} \right)}
      & {\rm if} \;\;\;\; l_{n} \ll \xi \ll a_{n} l_{n} & \\ \\
      - \nu^2(\epsilon_{Fn}) (4 \pi a_{n})^{-1/2}
      \exp {-\left[ (\xi - a_{n} l_{n})^2/4 a_{n} l_{n}^2 \right]}
      & {\rm if} \;\;\;\;
      \xi \simeq a_{n} l_{n} & \\
    \end {array}
  \right.,
\end {eqnarray}
\hfill
\begin{picture}(3.375,0)
  \put(0,0){\line(1,0){3.375}}
  \put(0,0){\line(0,-1){0.08}}
\end{picture}

\begin{multicols}{2}
\noindent where $ \xi = |y - y'| $, $ \epsilon_{Fn} = m^*
v_{Fn}^2/2 = k_{Fn}^2 / 2 m^* $, $ l_{n} $ is the electron
localization length in the $ n^{\rm th} $ subband, $ a_{n} = 2 \ln
{(8/\omega \tau_{n})} $, $ \tau_{n} = l_{n} / v_{Fn} $, and $
\omega = \epsilon - \epsilon' $. The dependence of $ F^{(1)}_{n} $
on $ \omega $ is weak, so that we can neglect it. The result that
is quoted above is valid for weak scattering $ k_{Fn} l_{n} \gg 1
$, and for a short range random potential $ \Lambda_{n} \ll l_{n}
$, where $ \Lambda_{n} $ is the correlation length of $ U_{n} $.

We are interested in small $ \epsilon_{Fn} $, since the transmission peaks
were observed when $ \epsilon_{Fn} $ approaches zero \cite {NBB96}.
However, when approaching the threshold, we are limited by the conditions
quoted above under which (\ref {F}) is valid. We shall now estimate how
close we can approach the threshold.

The localization length $ l_{n} $ is related to the one dimensional
transport scattering time via $ l_{n} = v_{Fn} \tau^{1D}_{n} $, where
$ \tau^{1D}_{n} $ is given by
\be
  \label {tau-1D}
  \frac {1}{\tau^{1D}_{n}} = \left( \frac {\pi}{2} \right)^{1/2}
  k_{Fn} \Lambda_{n} \frac {\langle U_{n}^{2} \rangle}{\epsilon_{Fn}}
  e^{-(k_{Fn} \Lambda_{n})^2/2}.
\ee
Here we considered $ U_{n}(y) $ to be Gaussian correlated with an amplitude
of $ \langle U_{n}^{2} \rangle $.
Since we have no information on the effective potential due to the roughness
of the QWR boundaries, we shall carry our estimates with $ U_{0} $, the random
potential for the unpatterned 2DEG, only.
The correlation length of $ U_{0} $ is given by the spacer width, which in
the particular experiment that we are interested in is $ 1050 \, {\rm  \AA} $.
The potential amplitude $ \langle U_{0}^{2} \rangle $ is related to the
two dimensional transport time via
\be
  \label {tau-2D}
  \frac {1}{\tau^{2D}} = \left( \frac {\pi}{8} \right)^{1/2}
  \frac {1}{k^{2D}_{F} \Lambda_{0}}
  \frac {\langle U_{0}^{2} \rangle}{\epsilon^{2D}_{F}},
\ee
where $ k^{2D}_{F} $ is the inverse Fermi wave length for the unpatterned
2DEG, that is related to the 2DEG density via
$ n^{2D} = (k^{2D}_{F})^2/2 \pi $, and to the two dimensional Fermi energy via
$ \epsilon^{2D}_{F} = (k^{2D}_{F})^2/2 m^* $. For
$ 1/ \tau^{2D} = 3 \times 10^{-2} \, {\rm meV} $ and
$ n^{2D} = 2.1 \times 10^{11} \, {\rm cm}^{-2} $ we find that
$ \langle U_{0}^{2} \rangle = 3.9 \, {\rm meV}^{2} $.

Returning to the conditions for the validity of (\ref {F}),
$ l_{n} \gg \Lambda_{n} $, and $ k_{Fn} l_{n} \gg 1 $, and replacing
$ U_{n} $ and $ \Lambda_{n} $, by $ U_{0} $ and $ \Lambda_{0} $, we find that
we can approach the threshold up to
$ \epsilon_{F}^{\rm th} = 1 \, {\rm meV} $. This corresponds to a change of
the magnetic field of $ \Delta B = 0.6 \, {\rm T} $.

\section {Calculation of $ \Pi $ and $ \epsilon^{-1} $}
\label {calculation}

Let us begin by dealing with the parallel case - that in which the phonons
propagate along the direction parallel to that of the QWR's
($ {\bf q} \parallel \hat {\bf y} $). In this case it suffices to use the
zeroth order wave functions in the matrix elements that appear in the
expression for $ \Pi $ (\ref {Pi-sq-wf}). Moreover, since the contribution of
terms with $ n \neq n' $ is much smaller than that of the terms with
$ n = n' $, due to the large energy denominator $ E_{n} - E_{n'} $, we keep
only the diagonal terms. Thus we have
\end{multicols}
\widetext

\noindent
\begin{picture}(3.375,0)
  \put(0,0){\line(1,0){3.375}}
  \put(3.375,0){\line(0,1){0.08}}
\end{picture}
\begin {equation}
  \label {Pi-sq-par}
  \Pi_{s,s'}({\bf q},\omega_{q}) \Bigl |_{{\bf q} =
             q \hat {\bf y}} =
  \frac {1}{L_{y}} \sum_{n} \int \! \! \int_{0}^{L_{y}} dy \, dy' e^{- i q (y-y')}
  \left \langle \sum_{\alpha,\alpha'}
    \chi^{*}_{n \alpha'}(y) \chi_{n \alpha}(y)
    \chi_{n \alpha'}(y')\chi^{*}_{n \alpha}(y')
    \frac {f(E^{0}_{n \alpha'})-f(E^{0}_{n \alpha})}
          {\omega_{q} - (\epsilon_{n \alpha} - \epsilon_{n \alpha'}) +
           i \delta}
  \right \rangle_{U}.
\end {equation}

Using the results presented in \cite {GDP83b}, we find
\begin {eqnarray}
  \label {Pi-sq-par1}
  \Pi_{s,s'}({\bf q},\omega_{q}) \Bigl |_{{\bf q} =
  q \hat {\bf y}} & = &
  \frac {1}{L_{y}} \sum_{n} \int \!\! \int_{0}^{L_{y}} dy \, dy' \,
  e^{- i q (y-y')}
  \int \!\! \int d \epsilon \, d \epsilon'
  \frac {\epsilon_{Fn}}{\sqrt {\epsilon \, \epsilon'}} \,
  \frac {f(E_{n}+\epsilon')-f(E_{n}+\epsilon)}
  {\omega_{q} - (\epsilon - \epsilon') + i \delta} \,
  F^{(1)}_{n}(y-y') \nonumber \\
  &
  = & \sum_{n} \bar {F}^{(1)}_{n}(q)
  \int \!\! \int d \epsilon \, d \epsilon'
  \frac {\epsilon_{Fn}}{\sqrt {\epsilon \, \epsilon'}} \,
  \frac {f(E_{n}+\epsilon')-f(E_{n}+\epsilon)}
  {\omega_{q} - (\epsilon - \epsilon') + i \delta}.
\end {eqnarray}
\begin{multicols}{2}
\noindent The function $ \bar {F}^{(1)}_{n}(q) $ is the Fourier
transform of $ F^{(1)}_{n}(y) $, and for $ q l_{n} \ll 1 $ it can
be approximated by $ \bar {F}^{(1)}_{n}(q) \Bigl |_{q l_{n} \ll 1}
\simeq (q l_{n})^2 l_{n} / v_{Fn}^2 $.

In order to conclude the calculation of
$ \Pi_{s,s'}({\bf q},\omega_{q}) \Bigl |_{{\bf q} = q \hat {\bf y}} $ we
still have to evaluate the integral over energies that appears in
expression (\ref {Pi-sq-par1})
\begin{eqnarray}
  \label {int-eps}
  \int \! \! \int d \epsilon \, d \epsilon'
  \frac {1}{\sqrt {\epsilon \, \epsilon'}} \,
  \frac {f(E_{n}+\epsilon')-f(E_{n}+\epsilon)}
  {\omega_{q} - (\epsilon - \epsilon') + i \delta} = \nonumber \\
  - \left(a + i \pi \frac {\omega_{q}}{\epsilon_{Fn}} \right),
\end{eqnarray}
where $ a $ is a numerical factor of the order of one. We have
used the following integral
\be
  \int d \epsilon \,
  \frac {1}{\sqrt {\epsilon (\epsilon - \omega_{q})}}
  \left[ f(E_{n} + \epsilon - \omega_{q}) - f(E_{n} + \epsilon) \right]
  \approx
  \frac {\omega_{q}}{\epsilon_{Fn}},
\ee
since due to the distribution functions $ \epsilon $ and
$ \epsilon - \omega_{q} $ must both be close to $ \epsilon_{Fn} $, and we
consider $ \omega_{q} \ll \epsilon_{Fn} $.

We finally obtain the following expression for $ \Pi_{s,s'} $ in the parallel
case
\begin{eqnarray}
  \label {Pi-sq-par2}
  \Pi_{s,s'}({\bf q},\omega) \Big |_{{\bf q} = q \hat {\bf y}} & =
  &
  - \sum_{n} \epsilon_{Fn} \bar {F}^{(1)}_{n}(q)
  \left (a + i \frac {\omega_{q}}{\epsilon_{Fn}} \right) \nonumber \\ &
  \equiv &
  \Pi^{\parallel}(q,\omega_{q}),
\end{eqnarray}
where the summation is only over occupied subbands.

In the parallel case $ \Pi_{s,s'}({\bf q},\omega) $ has no dependence on
$ s $ and $ s' $. Using this characteristic of $ \Pi $ we can now solve
Eq.\ (\ref {Eq-eps-sq1}) for the inverse dielectric function. We try a
solution of the form $ \delta_{s,s'} + V_{s}({\bf q}) g(q_{s'})/d $, substitute
it into Eq.\ (\ref {Eq-eps-sq1}) and find the function $ g(q_{s'}) $. For the
specific case of $ s = 0 $ and $ s' = 0 $ we obtain
\be
  \label {eps-00-par}
  \epsilon^{-1}_{0,0}({\bf q}, \omega_{q}) \Bigl |_{{\bf q} =
  q \hat {\bf y}} = 1 +
  \frac {V(q) \Pi^{\parallel}(q,\omega_{q})/d}
        {1 - V^{\parallel}(q)\Pi^{\parallel}(q,\omega_{q})/d},
\ee
where $ V^{\parallel}(q) = V(q) [1 + qd/\pi \ln ({d \gamma/2 \pi w})] $, and
$ \gamma \approx 1.8 $ is the Euler constant.

We now turn our attention to the perpendicular case. If we were to
keep only the zeroth order wave functions in the matrix elements that
appear in expression (\ref {Pi-sq-wf}), we would obtain that the
imaginary part of $ \Pi_{s,s'}({\bf q},\omega) \Bigl |_{{\bf q} = q
\hat {\bf x}} $ is zero, and thus no absorption occurs. We must
therefore include higher order corrections to the wave functions. In
fact, we must evaluate $ \Pi_{s,s'} $ up to the second order in $ W $
since the first order terms give zero contribution because they are
proportional to $ \langle W \rangle_{U} = 0 $. We shall, however,
consider the wave functions only up to the first order in $ W $, and
later on we shall explain why the second order correction to the wave
function is negligible in its contribution to $ \Pi_{s,s'} $.

The matrix element that appears in expression (\ref {Pi-sq-wf}) for
$ \Pi_{s,s'}({\bf q},\omega_{q}) \Bigl |_{{\bf q} = q \hat {\bf x}} $, up to
the first order in $ W $ is
\end{multicols}
\widetext

\noindent
\begin{picture}(3.375,0)
  \put(0,0){\line(1,0){3.375}}
  \put(3.375,0){\line(0,1){0.08}}
\end{picture}
\begin{equation}
  \label {exp-me1}
  \langle n' \alpha'|e^{i q_{s} x}|n \alpha \rangle =
  S_{n' \alpha';n \alpha}(\delta_{n',n} + i q_{s} x_{n' n}) +
  i q_{s}
  \sum_{m, \beta}
   \left[
    \frac {x_{m n} W_{m \beta;n \alpha}}{E^{0}_{n \alpha}-E^{0}_{m \beta}}
    \, S_{n' \alpha';m \beta} \,  x_{n' m} +
    \frac {x_{n' m} W_{n' \alpha';m \beta}}
          {E^{0}_{n' \alpha'}-E^{0}_{m \beta}}
    \, S_{m \beta;n \alpha} \, x_{m n}
   \right],
\end{equation}
\hfill
\begin{picture}(3.375,0)
  \put(0,0){\line(1,0){3.375}}
  \put(0,0){\line(0,-1){0.08}}
\end{picture}

\begin{multicols}{2}
\noindent where $ S_{n \alpha;n' \alpha'} =
        \int_{0}^{L_{y}} dy \, \chi^{*}_{n \alpha}(y) \chi_{n' \alpha'}(y) $
is the overlap integral of the electron wave functions along the wire
(it is not a delta function in the $ \alpha $ indices since the wave functions
$ \chi_{n \alpha}(y)$ and $ \chi_{n' \alpha'}(y) $ are defined by different
potentials).

The product of two matrix elements such as (\ref {exp-me1}) above should be
substituted into expression (\ref {Pi-sq-wf}) for $ \Pi_{s,s'} $. The
following terms out of this product will give zero contribution to
$ \Pi_{s,s'} $:
\begin {enumerate}
  \item Terms that are proportional to the diagonal (in subband indices)
        overlap integral
        $ S_{n' \alpha';n \alpha} \delta_{n',n} =
          \delta_{n',n} \delta_{\alpha', \alpha} $.
        This can be seen easily by substituting such a term into
        expression (\ref {Pi-sq-wf}).
  \item Terms that are proportional to $ W $. Since $ W(y) $ and $ U_{n}(y) $
        are uncorrelated, and $ \chi_{n}(y) $ are eigenfunctions of the
        Hamiltonian that contains the potential $ U_{n}(y) $, $ W(y) $
        and $ \chi_{n}(y) $ are also uncorrelated. Therefore, in the average
        that  appears in expression (\ref {Pi-sq-wf}) for $ \Pi_{s,s'} $
        we can separate the averaging over powers of $ W $ from that over
        the wave functions and the eigenenergies. Thus, if there is a term
        that is proportional to $ W $, after averaging it will be proportional
        to  $ \langle W(y) \rangle_{U} = 0 $.
\end {enumerate}

Expression (\ref {Pi-sq-wf}) includes the energy denominator
$ \omega_{q} - (E^{0}_{n \alpha} - E^{0}_{n' \alpha'}) + i \delta $.
When $ n \neq n' $ the energy difference
$ E^{0}_{n \alpha} - E^{0}_{n' \alpha'} $ is large, and
$ \omega_{q} + i \delta $
can be neglected compared to it (we have in mind the phonon frequencies for
$ \omega_{q} $ and these are much smaller than the subband energy difference).
Thus these terms will contribute only to the real part of $ \Pi_{s,s'} $,
and only terms with $ n=n' $ will contribute to the imaginary part.
There are additional energy denominators in the expression for $ \Pi_{s,s'} $
that come from the matrix elements. We shall keep only the largest terms,
those with the least number of large energy denominators. Had we kept the
second order correction to the wave function, it's contribution to
$ \Pi_{s,s'} $ would have been neglected at this point, since it includes too
many large energy denominators. We are left with the following expressions for
the real and imaginary parts of $ \Pi_{s,s'} $
\end{multicols}
\widetext
\be
  \label {Re-Pi-sq-per}
  {\rm Re} \, \Pi_{s,s'}({\bf q},\omega_{q}) \Bigl |_{{\bf q} =
  q \hat {\bf x}} =
  \frac {q_{s} q_{s'}}{L_{y}} \sum_{n,n'} |x_{n' n}|^2
  \left \langle
   \sum_{\alpha, \alpha'} |S_{n' \alpha';n \alpha}|^2
   \frac {f(E^{0}_{n' \alpha'}) - f(E^{0}_{n \alpha})}
         {E^{0}_{n' \alpha'} - E^{0}_{n,\alpha}}
  \right \rangle_{U},
\ee
and
\begin {eqnarray}
  \label {Im-Pi-sq-per}
  \lefteqn {{\rm Im} \, \Pi_{s,s'}({\bf q},\omega_{q}) \Bigl |_{{\bf q} =
  q \hat {\bf x}} =- \pi
  \frac {q_{s} q_{s'}}{L_{y}}} \\ \nonumber
  & \times & \sum_{n}
  \Biggl\langle
   \sum_{\alpha, \alpha'}
   \left| \sum_{m \beta} |x_{n m}|^2
     \left( S_{n \alpha';m \beta}
            \frac {W_{m \beta;n \alpha}}
                  {E^{0}_{n \alpha} - E^{0}_{m \beta}} +
            S_{m \beta;n \alpha}
            \frac {W_{n \alpha';m \beta}}
                  {E^{0}_{n \alpha'} - E^{0}_{m \beta}}
     \right) \right|^2
  \left[ f(E^{0}_{n \alpha'}) - f(E^{0}_{n \alpha}) \right]
  \delta (\epsilon_{n \alpha} - \epsilon_{n \alpha'} - \omega_{q})
  \Biggr\rangle_{U}.
\end {eqnarray}
\hfill
\begin{picture}(3.375,0)
  \put(0,0){\line(1,0){3.375}}
  \put(0,0){\line(0,-1){0.08}}
\end{picture}

\begin{multicols}{2}
In expression (\ref {Re-Pi-sq-per}) $ n \neq n' $ due to the
factor of $ |x_{n n'}|^2 $, therefore it can be simplified by
estimating the energy denominator to be $ E^{0}_{n' \alpha'} -
E^{0}_{n \alpha} \simeq E_{n'} - E_{n} $. We assume that the
electron wave functions in different subbands are uncorrelated.
Using the following definitions
\be
  \label {N}
  N_{n} = \frac {1}{L_{y}}
  \left \langle \sum_{\alpha} f(E^{0}_{n\alpha}) \right \rangle_{U},
\ee
the electron density in subband $ n $, and
\be
  \label {p}
  p_{n} = \sum_{n'} \frac {|x_{n n'}|^2}{E_{n} - E_{n'}},
\ee
the polarization related to subband $ n $, we can rewrite
(\ref {Re-Pi-sq-per}) as
\be
  \label {Re-Pi}
  {\rm Re} \, \Pi_{s,s'}({\bf q},\omega_{q}) \Bigl |_{{\bf q} =
  q \hat {\bf x}} =
  - 2 q_{s} q_{s'} \sum_{n} N_{n} p_{n}.
\ee

In expression (\ref {Im-Pi-sq-per}) the energy denominators
$ E^{0}_{n \alpha} - E^{0}_{m \beta} $, and
$ E^{0}_{n \alpha'} - E^{0}_{m \beta} $ can be approximated by
$ E_{n} - E_{m} $, since $ n \neq m $ due to the factor $ |x_{n m}|^2 $.
Furthermore, by separating the averaging over $ W $ from that over the wave
functions, and using the definitions of $ F^{(1)}_{n} $ (\ref {F}), and
$ p_{n} $ (\ref {p}), this term can be written as
\end{multicols}
\widetext

\noindent
\begin{picture}(3.375,0)
  \put(0,0){\line(1,0){3.375}}
  \put(3.375,0){\line(0,1){0.08}}
\end{picture}
\be
  \label {Im-Pi}
  {\rm Im} \, \Pi_{s,s'}({\bf q},\omega_{q}) \Bigl |_{{\bf q} =
  q \hat {\bf x}} =
  - 4 \pi \frac {q_{s} q_{s'}}{L_{y}} \omega_{q} \sum_{n} p_{n}^2
  \int \! \! \int_{0}^{L_{y}} dy \, dy'
  \langle W(y) W(y') \rangle_{U} F^{(1)}_{n}(|y - y'|).
\ee
The summation here is only over occupied subbands.

The correlator $ \langle W(y) W(y') \rangle_{U} $ is a function of $ y - y' $
only, and the length scale in which it changes is
$ \Lambda_{n} $, while the length scale in which $ F^{(1)}_{n}(|y - y'|) $
changes is $ l_{n} $. Since we consider the case in which
$ \Lambda_{n} \ll l_{n} $ we may write
\be
  \int \! \! \int_{0}^{L_{y}} dy \, dy' \langle W(y) W(y') \rangle_{U}
            F^{(1)}_{n}(|y - y'|) =
            L_{y} F^{(1)}_{n}(0) \int_{-\infty}^{\infty} dy
            \langle W(y) W(0) \rangle_{U} \equiv
  L_{y} F^{(1)}_{n}(0) \langle W^2 \rangle_{q=0} .
\ee Thus we can finally write
\be
  \label {Pi-sq-per1}
  \Pi_{s,s'}({\bf q},\omega_{q})|_{{\bf q} = q \hat {\bf x}} =
  - 2 q_{s} q_{s'} \sum_{n} \left(
  N_{n} p_{n} +
  \frac {i}{3 \pi} m^* p_{n}^2 \frac {\omega_{q}}{\epsilon_{Fn}}
  \langle W^2 \rangle_{q=0} \right) \equiv
  \frac {q_{s} q_{s'}}{q^2} \Pi^{\perp}(q, \omega_{q}).
\ee \hfill
\begin{picture}(3.375,0)
  \put(0,0){\line(1,0){3.375}}
  \put(0,0){\line(0,-1){0.08}}
\end{picture}

\begin{multicols}{2}
In the perpendicular case $ \Pi_{s,s'}({\bf q},\omega_{q}) \propto
  q_{s} q_{s'} $.
Using this characteristic of $ \Pi $ we substitute a solution of the form
$ \delta_{s,s'} + q_{s} V_{s}({\bf q}) g(q_{s'}) $ for
$ \epsilon^{-1}_{s,s'} $ into Eq.\ (\ref {Eq-eps-sq1}),
and find $ g(q_{s'}) $. For the specific case of $ s=0 $ and $ s'=0 $ we
find
\be
  \label {eps-00-per}
  \epsilon^{-1}_{0,0}({\bf q},\omega_{q}) \Bigl |_{{\bf q} =
  q \hat {\bf x}} = 1 +
  \frac {V(q) \Pi^{\perp}(q,\omega_{q}) / d}
        {1 - V^{\perp}(q) \Pi^{\perp}(q,\omega_{q}) / d},
\ee
where $ V^{\perp}(q) = V(q) (1 + d/ 2 \pi q w^2) $.

The SAW absorption, and its relative change of velocity, can now be found
by substituting expressions (\ref {eps-00-par}) and (\ref {eps-00-per})
for $ \epsilon^{-1}_{0,0} $ in the parallel and perpendicular cases, into
expressions (\ref {Gamma}) and (\ref {dv/v}).

\section {Discussion}
\label {discussion}

Let us begin by discussing expressions (\ref {eps-00-par}) and
(\ref {eps-00-per}) for $ \epsilon^{-1}_{0,0} $. From these expressions one
can see that we could not have replaced $ \sigma_{\rm dc} $ by
$ \sigma_{\rm ac} $ in the ``usual'' expression for the attenuation
$ \Gamma $ (\ref {Gamma-dc}). The numerators in these expressions describe
the unscreened interaction, while the denominators describe the screening.
This can be seen by expanding the expressions for $ \epsilon^{-1}_{0,0} $ in
the electron charge. The combinations of $ V \Pi $ that enter the screening
and the interaction are different, therefore the attenuation cannot be
described by (\ref {Gamma-dc}) with $ \sigma_{\rm dc} $ replaced by
$ \sigma_{\rm ac} $.

One can see from expressions (\ref {Pi-sq-par}) and (\ref {Pi-sq-per1}) that
in both cases $ {\rm Im} \Pi \ll {\rm Re} \Pi $ due to the factor of
$ \omega_{q}/ \epsilon_{Fn} $, thus the screening is
dominated by $ {\rm Re} \Pi $. The ratio $ \Pi/d $ that appears in the
expressions for $ \epsilon^{-1}_{0,0} $ is the effective two dimensional
polarization of the patterned 2DEG.

In the parallel case, as $ \epsilon_{Fn} $ increases, that is, as we move
away from the threshold, the localization $ l_{n} $ increases, and so does
the screening. This can be understood in the following manner: When the
electron energy increases, the random potential seems weaker, and therefore
the localization length grows. When the localization length increases, the
electrons can move more freely, and the thus the screening is enhanced.

In the perpendicular case the mechanism through which the screening increases
as $ \epsilon_{Fn} $ increases is different. As the bottom of the subband is
lowered, the electron number in the QWR increases, and therefore the screening
increases. One can see that as the wire width $ w $ decreases towards zero
$ \epsilon^{-1}_{0,0} $ approaches one, so there is no screening. This is
due to the fact that $ x_{n n'} $ that appears in $ p_{n} $ (\ref {p}), which
in turn appears in the expression for $ \Pi^{\perp} $
(\ref {Pi-sq-per1}), is proportional to $ w $.

We find that in both the parallel and the perpendicular geometries
$ {\rm Im} \, \epsilon^{-1}_{0,0} \propto 1 / \epsilon_{Fn} $ (in the parallel
case it is due to $ \bar {F}^{(1)}_{n}(q) \propto 1/v_{Fn}^2 $), and therefore
the absorption should show a structure that is related to the crossing of
the Fermi energy by the bottom of the subbands in {\em both} geometries,
unlike what was seen in the experiment \cite {NBB96}.

We can now make some numerical estimates of the magnitude of the phonon
absorption per unit length $ \Gamma $, and the phonon change in velocity
$ \Delta v / v $, for both the parallel and the perpendicular cases.
We consider an array of wires of period $ d = 10^4 \, {\rm \AA} $ and of width
$ w = 5000 \, {\rm \AA} $ manufactured in GaAs, as in Ref.\ \cite {NBB96}. We
take the localization length $ l_{n} \simeq 1 \, \mu {\rm m} $, in accordance
with experimental work done in quantum wires \cite {TRS89}. We find that in
both cases $ \Gamma \simeq 10^{-4} \, {\rm cm^{-1}} $ and
$ \Delta v / v \simeq - (10^{-5} - 10^{-4}) $.

The fact that our results do not fully agree with the experimental findings
raises the question whether the magnetic field should not be considered more
accurately. We considered it only as a mechanism for moving the bottom of
the energy subbands, and yet the magnetic field should affect both the
electron wave functions, and the localization length (that should increase
when a magnetic field is turned on).

\acknowledgements

We would like to thank D. Khmelnitskii for helpful discussions. One of us
(Y. L.) would like to thank G. R. Nash and S. J. Bending for the discussions
of the experimental aspects, and the British Council in Tel Aviv for its
support.

\end{multicols}


\begin{thebibliography}{10}

\bibitem{WKW86}
A. Wixforth, J.~P. Kotthaus, and G. Weimann, Phys. Rev. Lett. {\bf
56},  2104
  (1986).

\bibitem{WSW89}
A. Wixforth {\it et~al.}, Phys. Rev. B {\bf 40},  7874  (1989).

\bibitem{WPR90}
R.~L. Willet {\it et~al.}, Phys. Rev. Lett. {\bf 65},  112
(1990).

\bibitem{WRP93}
R.~L. Willet {\it et~al.}, Phys. Rev. B {\bf 47},  7344  (1993).

\bibitem{WRW93}
R.~L. Willet, R.~R. Ruel, K.~W. West, and L.~N. Pfeiffer, Phys.
Rev. Lett. {\bf
  71},  3846  (1993).

\bibitem{WWP97}
R.~L. Willet, K.~W. West, and L.~N. Pfeiffer, Phys. Rev. Lett.
{\bf 78},  4478
  (1997).

\bibitem{LEM98}
Y. Levinson, O. Entin-Wohlman, A.~D. Mirlin, and P. W\"olfle,
Phys. Rev. B {\bf
  58},  7113  (1998).

\bibitem{MWL98}
A.~D. Mirlin, P. W\"olfle, Y. Levinson, and O. Entin-Wohlman,
Phys. Rev. Lett.
  {\bf 81},  1070  (1998).

\bibitem{OSH98}
F. von Oppen, A. Stern, and B.~I. Halperin, Phys. Rev. Lett. {\bf
80},  4494
  (1998).

\bibitem{NBB96}
G.~R. Nash {\it et~al.}, Phys. Rev. B {\bf 54},  R8337  (1996).

\bibitem{KLE96}
A. Kn\"abchen, Y. Levinson, and O. Entin-Wohlman, Phys. Rev. B
{\bf 54},  10696
   (1996).

\bibitem{LeR85}
P.~A. Lee and T.~V. Ramakrishnan, Rev. Mod. Phys. {\bf 57},  287
(1985).

\bibitem{Mah}
G.~D. Mahan, {\em Many particle physics} (Plenum Press, New York,
1990).

\bibitem{LiP}
E.~M. Lifshitz and L.~P. Pitaevskii, {\em Physical Kinetics},
Vol.~10 of {\em
  Course of Theoretical Physics} (Pergamon Press, Oxford, 1981).

\bibitem{GDP83b}
L.~P. Gor'kov, O.~N. Dorokhov, and F.~V. Prigara, Sov. Phys. JETP
{\bf 58},
  852  (1983).

\bibitem{TRS89}
T.~J. Thornton, M.~L. Roukes, A. Scherer, and B.~P.~V. de~Gaag,
Phys. Rev.
  Lett. {\bf 63},  2128  (1989).

\end{thebibliography}
\end {document}